\documentclass[11pt]{article}
\usepackage{bbold}
\usepackage{amsmath}
\begin{document}
\date{}
\title{Time Coordinates and Clocks: Einstein's Struggle}
\author{Dennis Dieks \\History and Philosophy of Science\\
Utrecht University\\ d.dieks@uu.nl}
\maketitle
\begin{abstract}
In his Autobiographical Notes, Einstein mentioned that on his road to the final theory of general relativity it was a major difficulty to accustom himself to the idea that coordinates need not  possess an immediate physical meaning in terms of lengths and times. This appears strange: that coordinates are conventional markers of events seems an obvious fact, already familiar from pre-relativistic physics. In this paper we explore the background of Einstein’s difficulties, going from his 1905 paper on special relativity, through his 1907 and 1911 papers on the consequences of the equivalence principle, to the 1916 review paper on the general theory. As we shall argue, Einstein's problems were intimately connected to his early methodology, in which clarity achieved by concrete physical pictures played an essential role; and to the related fact that on his route to the general theory he focused on special situations that were easily accessible to physical intuition.
The details of this background of Einstein's early reasoning have not always been sufficiently appreciated in modern commentaries. As we shall see, this has led to erroneous judgments about the status and validity of some of the early relativistic derivations by Einstein and others, in particular concerning the gravitational redshift.
 \end{abstract}

\section{Introduction}\label{intro}

In his Autobiographical Notes \cite[pp.\ 66--67]{einsteinauto}, Einstein remarks that during his work on the general theory of relativity an important obstacle for progress was his own reluctance to abandon the idea that coordinates should possess an immediate physical  meaning in terms of measurable distances and times.  At first sight, this strikes one as strange: that coordinates are conventional and can be chosen arbitrarily as long as they uniquely identify physical events seems an obvious fact already familiar from pre-relativistic physics.  In this paper we explore the background of Einstein’s difficulties, going from his 1905 paper on special relativity \cite{einstein1905}, through his 1907 and 1911 papers \cite{einstein1907,einstein1911} on the consequences of the principle of equivalence, to the definitive 1916 review paper on the general theory \cite{einstein1916} and some later developments.

An essential ingredient of Einstein's 1905 paper was the analysis of physical time, in particular the notion of distant simultaneity. The purpose of this analysis was to become as clear and concrete as possible about the role of simultaneity in concrete physical situations, in order to show that what classical physics assumed about time and simultaneity was partly unfounded. For this reason Einstein discussed an almost everyday synchronization procedure, via light signals exchanged between (ideal) macroscopic clocks. Clearly, simultaneity specified this way does not consist in the equality of arbitrarily chosen time coordinates, but is a concrete physical notion relating to the indications of clocks. 

We encounter the same strategy of arguing on the basis of concrete physical situations that are not too different from familiar cases in Einstein’s 1907 and 1911 papers. Here Einstein discusses the principle of equivalence, and derives the gravitational redshift as one of its consequences.  In his 1911 article Einstein focuses on the comparison of similarly constructed clocks placed at different positions in a homogeneous gravitational field. As Einstein argues on the basis of symmetry considerations, from the point of view of local observers, near to the clocks, such clocks must all behave identically and tick at the same rate. However, Einstein continues, in addition to these local times we should also introduce a \emph{global} time, in order to compare processes at different positions. This alternative time is also introduced via clocks and signals that connect them, via procedures suggested by the physical characteristics of the global situation. It then turns out that the original (``local'') clocks and the new``global'' ones do not agree in their indications. 

This introduction of a global time is not a matter of introducing a conventionally chosen coordinate. Rather than presenting his global time as an arbitrary marker, Einstein explains that it is a physical quantity that is measured by actual clocks, synchronized via a physically justified procedure. The same line of thought occurs in the beginning of the 1916 general relativity paper, in which Einstein discusses temporal relations on a rotating disk. Here he emphasizes that clocks at the periphery of the disk \emph{really} run slower than a clock at the center, again via the introduction of a physically significant global time that is measured by a set of suitable clocks. The notion that there exists a ``real'' global time which is represented by the indications of a privileged set of clocks apparently had a strong hold on Einstein, even at the time when he was presenting an overview of his general theory of relativity.

It is clear why Einstein felt that the introduction of a global notion of time was necessary: we need to compare and connect local situations in order to be able to make predictions concerning processes that are spatially extended. A prime example is the propagation of light: how long does it take for a light signal to go from $A$ to $B$, and how do the frequencies at emission and reception compare? Another and related example is the gravitational deflection of light that grazes the Sun (the best-remembered subject of Einstein's 1911 paper). Einstein's theoretical derivations of these effects, in 1907, 1911 and 1916, make essential use of the differences between the indications of local clocks in a gravitational field and clocks that measure global time. As we shall argue, not all aspects of these derivations have always been represented correctly in modern commentaries, which has led to a number of unjustified criticisms of Einstein's (and others') early arguments, in particular concerning the gravitational redshift. Consideration of this episode in the history of relativity theory thus opens up both a perspective on Einstein’s early methodological thinking and on the status of a number of early relativistic derivations.

Present-day relativists are well aware that there is no pre-determined metrical structure in general relativity; that the theory is ``background-independent''. This background independence has the consequence that the preferred global time coordinates that were assumed to exist by Einstein in his early work, cannot be expected to exist in general. The transition from Einstein's global time as a privileged coordinate, measurable by physical clocks, to the modern concept of arbitrarily chosen temporal coordinates is closely connected to the transition from special relativity, with its \emph{a priori} symmetries, to general relativity with its lack of a pre-given metrical structure. The account given in this paper is intended to contribute to understanding why Einstein only gradually became comfortable with the idea that space-time coordinates cannot always be given a direct physical and intuitive meaning.

\section{Einstein's 1905 paper}\label{SRT}

In the Introduction of his 1905 paper Einstein famously declared \cite[p.\ 38]{einsteinlorentz}: ``The theory to be developed is based---like all electrodynamics---on the kinematics of the rigid body, since the assertions of any such theory have to do with the relationships between rigid bodies (systems of coordinates), clocks, and electromagnetic processes.'' On the same page he continues (section 1 of the paper): ``If we wish to describe the \emph{motion} of a material point, we give the values of its coordinates as a function of the time.  Now we must bear carefully in mind that a mathematical description of this kind has no physical meaning unless we are quite clear as to what we understand by `time'.'' 

The necessary clarification of the meaning of time is given by Einstein in two steps: first we need to avail ourselves of clocks, calibrated in such a way that they locally indicate the familiar time of mechanics (so that ``the equations of Newtonian mechanics hold good to the first approximation'' \cite[p.\ 38]{einsteinlorentz}).   Now, if we have two such clocks, one at position A, and another of exactly the same construction at B, we have defined an ``A time'' and a ``B time'', but not yet a common time for A and B. In order to establish this common, global time Einstein in his second step introduces the celebrated special relativistic synchronization rule \cite[p.\ 894]{einstein1905}, \cite[p.\ 40]{einsteinlorentz}: ``Let a ray of light start at the `A time' $t_A$ from A towards B, let it at the `B time' $t_B$ be reflected at B in the direction of A, and arrive again at A at the `A time' $t^{\prime}_A$. By definition, the two clocks synchronize if $t_B - t_A = t^{\prime}_A - t_B$.''

This procedure settles the physical content of synchronicity of stationary clocks at different places. We can now imagine an inertial system filled with such clocks, all synchronized with one standard clock (for example, the clock at the origin of the coordinate system). The times thus assigned plus the spatial coordinates provided by measuring rods give a concrete physical meaning to the spatiotemporal description of processes.\footnote{Einstein's 1905 explanation of the physical meaning of space and time coordinates is sometimes interpreted as being part and parcel of an operationalist philosophy of science, according to which the meaning of concepts is nothing but what is \emph{defined} (in the mathematical-logical sense) by a procedure with measuring instruments. However, the 1905 paper is obviously meant to make the introduction of new physics, with new concepts, acceptable, and is not intended to defend some particular position in the philosophy of science---operationalism is not the issue at stake here. Accordingly, the synchronization procedure proposed by Einstein should not be seen as a strict \emph{definition} of simultaneity in the logical sense, but rather as a concrete implementation of the simultaneity relation that is close to everyday experience and physical intuition (cf. \cite{dieks}). To mention just one relevant argument, special relativistic physics, and relativistic space and time, will clearly still be well-defined in situations in which there can exist no clocks or measuring rods. However, this question and the related one whether the value of the one-way velocity of light is purely conventional are tangential to the theme of this article and will not be discussed here.} 

In his 1905 paper Einstein thus introduces a local and a global time, both measurable by clocks (in macroscopic laboratory contexts). In fact, \emph{one} set of clocks suffices: synchronization of the local clocks makes them into indicators of global time as well. This reflects a peculiarity of special relativistic inertial frames, which is due to the symmetry  of Minkowski spacetime. As we shall see in the next two sections, \emph{accelerated} frames in Minkowski spacetime, and \emph{a fortiori }frames in general relativistic spacetimes, in general do not admit such a simple connection between local and global time.

\section{The 1907 paper: the principle of equivalence}\label{1907}

In 1907 Einstein wrote an extensive review of his recently proposed relativity theory: ``On the relativity principle and the conclusions drawn from it'' \cite{einstein1907}. After having reviewed known territory, in the final part of the paper Einstein poses the question whether the relativity principle might be extended so as to apply also to \emph{accelerated} frames of reference. In order to answer this question Einstein considers two reference frames: an inertial system $\Sigma_2$, which is located in a homogeneous gravitational field with a free-fall acceleration of $-\gamma$ in the direction of the $X$-axis, and another frame $\Sigma_1$ that is uniformly accelerated in empty space with an acceleration $\gamma$ along the $X$-axis. Now, Einstein states, as far as we know the physical laws in $\Sigma_1$ do not differ from those in $\Sigma_2$: all bodies are equally accelerated in a gravitational field, so that we see exactly the same motions in $\Sigma_2$ as relative to the accelerated frame $\Sigma_1$. Einstein concludes that in our present state of knowledge we have no reason to assume that the systems $\Sigma_1$ and $\Sigma_2$ differ in any respect. In other words, we may suppose that there exists a physical equivalence between an inertial system with a gravitational field and an accelerated frame of reference. This is the first appearance in the literature of the celebrated principle of equivalence.\footnote{Einstein´s text may suggest that the principle of equivalence \emph{follows} from the empirical fact that all bodies are equally accelerated in a homogeneous gravitational field. That is not the case, of course; the \emph{complete} equivalence posited by Einstein constitutes a hypothesis that goes far beyond the direct empirical evidence of equality of accelerated motions.} 

As Einstein emphasizes, the equivalence principle has great heuristic value, because it enables us to make predictions about what will happen in a gravitational field: we can replace a constant and homogeneous gravitational field by a uniformly accelerated frame of reference, which is accessible to theoretical treatment via the special theory of relativity.

In order to elaborate this latter point, Einstein invites us to consider the space-time relations between an inertial system $S$ and a system $\Sigma$ that is uniformly accelerated along  the $X$-axis of $S$, with acceleration $\gamma$. Both systems are supposed to be  equipped with measuring rods and clocks of exactly the same construction. At $S$-time $t=0$ the frames $S$ and $\Sigma$ are assumed to coincide and to be instantaneously at rest with respect to each other. At that point every clock in $\Sigma$ is adjusted so as to indicate the same time as the corresponding clock in $S$.  The time indicated by the individual clocks in $\Sigma$, after this initial synchronization, is called the ``local time'' $\sigma$ of $\Sigma$. As Einstein observes, in terms of this local time the description of physical processes will be locally everywhere the same in system $\Sigma$, i.e.\ independent of spatial position (indeed, this is clear when we realize that all points in $\Sigma$ move in exactly the same way against the homogeneous background of Minkowski spacetime). 

However, Einstein continues, we cannot simply consider $\sigma$ as giving us ``the time'' of system $\Sigma$. This is because the accelerating clocks of $\Sigma$, indicating $\sigma$, will not remain synchronous with respect to each other: after their initial synchronization at $t=0$ two clocks that show the same value of $\sigma$ will no longer satisfy the simultaneity criterion (as introduced in the 1905 paper). This can easily be seen in the following way: since all clocks in $\Sigma$ execute exactly the same accelerated motion, with respect to $S$, they will remain synchronized as judged from $S$ (in which they were synchronized at $t=0$)---indeed, at any instant of $S$, all $\Sigma$-clocks will have ticked away the same amount of time since t=0, as judged from $S$. But this synchronicity with respect to $S$ entails that the $\Sigma$-clocks can no longer be synchronized as viewed from an inertial system $S^{\prime}$ that is instantaneously at rest relative to $\Sigma$ at any later time: at any moment of $S$-time later than t=0, $\Sigma$ and therefore also $S^{\prime}$, will have a non-zero velocity with respect to $S$ so that the simultaneity relations in $S$ and $S^{\prime}$ differ, as we know from special relativity.

In addition to the local time $\sigma$ Einstein therefore defines a ``global time'': the global time $\tau$ of an event in $\Sigma$ is the time indicated by a clock at the origin of $\Sigma$ at the instant that is simultaneous with the event in question, according to the simultaneity of the instantaneously comoving inertial system $S^\prime$. As we have just seen, it follows that the local time $\sigma$ and this global time $\tau$ in $\Sigma$ are different. The global time $\tau$ coincides with the time of the instantaneously comoving frame $S^\prime$ (if the clocks at their common origin are set to agree).

The quantitative relation between $\sigma$ and $\tau$ can be determined on the basis of the special theory of relativity. Two events that take place at positions $x_1$ and $x_2$ and times $t_1$ and $t_2$ of $S$, respectively, are simultaneous with respect to $S^\prime$ if $t_1 - x_1 v/c^2 = t_2 - x_2 v/c^2 $, in which $v$ is the speed of $S^\prime$ with respect to $S$. In the case of a small time difference with $t=0$, and a small velocity $v$, we have in first approximation $x_2 - x_1 = x^{\prime}_2 - x^\prime_1 = \xi_2 -\xi_1$,\footnote{As Einstein points out, only speeds and not accelerations have a systematic effect on lengths.} with $\xi$ the coordinate in system $\Sigma$ along the common $X$-axis. Moreover, we have in this approximation $t_1 = \sigma_1$, $t_2 = \sigma_2$, and $v = \gamma \tau$.\footnote{In his 1908 brief correction and addition to the 1907 paper \cite{einstein1908} Einstein points out that in a rigorous treatment the constant acceleration $\gamma$ should be defined with respect to the instantaneously comoving system $S^\prime$, so that it is not constant with respect to $S$.}

It follows that $\sigma_2 - \sigma_1 = (\xi_2 - \xi_1) (\gamma \tau) /c^2 $. When we take the origin of the coordinate system as the place of the first event, so that $\sigma_1 = \tau$ and $\xi_1 = 0$, we obtain: 
\begin{equation}
\sigma = \tau ( 1 + \frac{\gamma \xi}{c^2} ).\label{grav}
\end{equation} 

Now, according to the equivalence principle, this equation should also hold in a system in which there is a homogeneous gravitational field. In this case we can replace $\gamma \xi$ by the gravitational potential $\Phi$, so that we obtain 
\begin{equation}\label{redshift1907}
\sigma = \tau ( 1 + \frac{\Phi}{c^2} ).
\end{equation}            

So, summarizing, there are two time systems in $\Sigma$: the local time $\sigma$ and the global time $\tau$. Both are indicated by sets of clocks: the local time by the clocks that were initially synchronized with the clocks in $S$ and then move along with $\Sigma$ without further interference; and the successive sets of synchronized clocks in the instantaneously comoving frames $S^\prime$, with the understanding that the clocks at the origin of the comoving frame is set to show the same time as the $\sigma$-clock at the origin of $\Sigma$. 

When we are interested in local processes and local measurements, it is natural to use the $\sigma$-clocks, which are everywhere in the same physical state and run at the same local rate. However, when we describe extended processes we need a notion of simultaneity to formulate the pertinent physical laws (for example those governing the propagation of light signals), and in this case the $\tau$-clocks will be appropriate because equality of $\tau$ values signifies simultaneity in the standard sense.

It follows from Eq.\ (\ref{grav}) that clocks in a gravitational field, at a position with gravitational potential $\Phi$, indicate time values that differ from the time of the clock at the origin (where $\Phi = 0$) by a factor $(1 + \Phi / c^2)$. It is important to note, as Einstein stresses, that this difference has an immediate experimental significance: a stationary observer in $\Sigma$ who looks at the two clocks will  observe that they tick at different rates, given by exactly this factor. This is because the time intervals needed by the light to reach the observer are independent of $\tau$ \cite[p.\ 458]{einstein1908}---there is no time-dependence in the factor by which $\tau$ differs from $\sigma$, according to formula (\ref{grav}). This constitutes the background for the statement that clocks at a position with a higher gravitational potential \emph{really} run faster: the global time system $\tau$ makes it possible to compare the local rates in a way that is directly empirically verifiable.  

An example of the influence of gravity on the rate of clocks is provided by the behavior of atoms and molecules that emit spectral lines: these atoms and molecules can be considered to be clocks with a highly stable frequency. From the just-given argument it can be concluded that spectral light coming from the surface of the Sun will arrive on Earth with a frequency that is slightly shifted to the red end of the spectrum.

Another prediction is that electromagnetic radiation will be bent in a gravitational field \cite[p.\ 461]{einstein1907}. The derivation of this prediction in the 1908 paper is rather cumbersome, though---Einstein comes back to the topic in his 1911 paper, to which we turn now.

\section{The 1911 paper on gravity, time and light}\label{1911} 

In the Introduction of his 1911 paper Einstein states to return to a number of questions already treated in his 1907 paper, partly because his earlier discussion no longer satisfies him, but primarily because he now realizes that one of his earlier predictions can be tested experimentally. In fact, star light grazing the Sun will be bent by the Sun's gravitational field, and this will lead to a detectable apparent shift in the positions of fixed stars that appear near to the Sun in the sky. The declared objective of the paper is to explain this and other gravitational phenomena by very elementary considerations, so as to make the basic assumptions and arguments of the theory easily understandable.

Einstein again bases his considerations on the principle of equivalence, now making it more explicit than before that this principle represents a fundamental new hypothesis about the nature of gravitation and is meant to extend to all possible physical phenomena. Then he introduces a simple thought experiment in which the equivalence principle is used to derive several results concerning gravity; we shall only discuss the parts pertinent to the behavior of clocks in a gravitational field.    

Let there be two bodies $A$ and $B$,\footnote{We change the notation from Einstein's own in order to avoid confusion with symbols used in the previous section.} in a system of reference $K$ in which there exists a homogeneous gravitational field. Both systems, assumed to be infinitely small, are positioned on the $z$-axis of $K$; the acceleration due to gravity is $-\gamma$ (directed downward along the $z$-axis) and system $B$ is located higher in the field, at a distance $h$ from $A$, so that the gravitational potential at $B$ is greater than at $A$, $\Phi(B) - \Phi(A) = \gamma h $. The system $B$ emits a quantity of electromagnetic radiation in the direction of $A$, and we are interested in the influence of gravity on the properties and propagation of this radiation.

By virtue of the principle of equivalence we can replace system $K$ with a system $K^\prime$ that possesses a constant acceleration $\gamma$ in the positive $z$-direction and in which there is no gravitational field. In order to have a situation that is equivalent to the original one we have to assume that $A$ and $B$ are located at fixed positions on the $z^\prime$-axis, with the constant mutual distance $h$. Finally, let $K_0$ be an inertial system that at the moment of the emission of the radiation is instantaneously at rest with respect to $K^\prime$. 

When we describe the process of the emission, propagation and reception of the radiation from system $K_0$, we have that $B$ has no velocity relative to $K_0$ when the radiation is emitted, that the radiation then takes a time  $h/c$ to arrive at $A$ (in first approximation), and that $A$ possesses the approximate speed $ (h \gamma)/c = v$ when the radiation arrives. Now, Einstein notes, if the radiation had the frequency $\nu_2$ when it was emitted by $B$, as measured by a standard clock positioned at $B$, the radiation received in $A$ will have a different frequency $\nu_1$ as measured by a clock of the same construction comoving with $A$. Indeed, when the signal arrives at $A$, $A$ (and its clock) will possess a speed $v$ relative to $K_0$, so that there will be a change in measured frequency on account of the Doppler effect. The relation between $\nu_1$ and $\nu_2$ is given by the Doppler formula
\begin{equation}\label{doppler}
\nu_1 = \nu_2 (1 + \frac{\gamma h}{c^2}).
\end{equation} 

According to the equivalence principle this result also holds for the system $K$ in which there is a gravitational field. That means that light emitted at a higher value of te gravitational potential will arrive at positions with a lower potential with a higher frequency (as measured by a local clock). Rewriting Eq.\ (\ref{doppler}) for this case, we find 
\begin{equation}\label{redshift1911}
\nu_1 = \nu_2 (1 + \frac{\Phi}{c^2}),
\end{equation}   
where $\Phi$ is the gravitational potential at $B$ and the value of the potential at $A$ has been set to $0$. This is the same gravitational redshift formula as derived in the 1907 paper, Eq.\ (\ref{redshift1907}).

The 1911 derivation of the redshift formula may seem very different from that in the 1907 paper: in 1907 Einstein emphasized the necessity of a global time, whereas the 1911 derivation appears to involve only local times, measured by clocks of the same kind positioned at different positions in the gravitational field. However, that impression would be deceptive. First, the application of the Doppler formula presupposes that there is a global time system by means of which we can compare the frequencies at $A$ and $B$. This global time is provided by inertial system $K_0$ with its standard simultaneity. As we have seen in section \ref{1907}, however, the clock that is stationary at $A$ will not agree with this $K_0$ time, because during the transmission process it obtains a velocity relative to $K_0$. Conversely, if we want to describe the process from the viewpoint of a stationary observer at $A$, with the clock at $A$ as his fiducial clock, we should introduce a global time via the standard synchronization procedure from $A$. This is in essence the same procedure as in section \ref{1907} and leads to the same global time that was introduced there. Of course, compared to this global time the local clock in $B$ will be out of step, and formula (\ref{grav}) applies.      

Einstein discusses the situation as follows \cite[pp.\ 905--906; pp.\ 105--106 in the English translation]{einstein1911}:
\begin{quote}
On superficial consideration equation (\ref{redshift1911}) seems to assert an
absurdity. If there is constant transmission of light from $B$ to $A$, how can any other
number of periods per second arrive at $A$ than is emitted from $B$? But the answer is simple. We cannot regard $\nu_2$ or respectively $\nu_1$ simply as frequencies (as the number of periods per second) since we have not yet determined a time in system $K$. What
$\nu_2$ denotes is the number of periods per second with reference to the time-unit of the
clock $U$ at $B$, while $\nu_1$ denotes the number of periods per second with reference to
the identical clock at $A$. Nothing compels us to assume that the clocks $U$ in different
gravitation potentials must be regarded as going at the same rate. On the contrary,
we must certainly define the time in $K$ in such a way that the number of wave crests
and troughs between $B$ and $A$ is independent of the absolute value of time: for the
process under observation is by nature a stationary one. ... Therefore the two clocks at $A$ and $B$ do not both give the ``time''
correctly. If we measure time at $A$ with the clock $U$, then we must measure time
at $B$ with a clock which goes $1 + \Phi/c^2$
times more slowly than the clock $U$ when
compared with $U$ at one at the same place. For when measured by such a clock,
the frequency of the light-ray which is considered above is at its emission from $B$ given by $\nu_2 (1 + \Phi/c^2)$,
and is therefore, by (\ref{redshift1911}), equal to the frequency $\nu_1$ of the same light-ray on its arrival
at $A$.
\end{quote} 

Einstein here explicitly introduces a global time that differs from the local time; the global time corresponds to the time $\tau$ defined in the 1907 article, while the clocks $U$ correspond to the local time $\sigma$. As in the 1905 and 1907 articles, both the local and global times are assumed to be directly measured by sets of clocks.  In the 1907 article this material implementation of global time was realized by standard clocks in the instantaneously comoving inertial systems like $S^\prime$, whereas in the 1911 paper the clocks indicating global time are introduced directly, via the rule that they be constructed thus that they tick $1 + \Phi/c^2$ times more slowly than local clocks at the same location.\footnote{In the Collected Papers of Einstein, Volume 3, the editors comment that Einstein's train of thought in the 1911 paper is quite different from the one in 1907 \cite[p.\ 497]{einsteincoll3}. In particular, they claim that the slow clocks of the 1911 paper played no role in the 1907 article, and refer to \cite[pp.\ 198--199]{pais} for a further analysis of the significance of these clocks. In the indicated passage Pais suggests that clocks produced at $A$ will automatically run slower by the factor $1 + \Phi/c^2$ when transported to $B$. This seems a misunderstanding: clocks transported to other positions will tick at the rate of \emph{local} clocks, whereas the slower clocks were introduced by Einstein to measure \emph{global} time. In fact, the slow clocks \emph{did} occur in the 1907 paper, although in the guise of standard clocks in instantaneously comoving frames.}

The 1911 paper ends with a calculation of the bending of light in a gravitational field, and is most famous for this prediction. This calculation is based on the observation that the velocity of light, measured in global time, will not be constant but will vary with the gravitational potential according to
\begin{equation}
c = c_0 (1 + \frac{\Phi}{c^2}),
\end{equation} 
where $c_0$ is the value at the origin (where $\Phi = 0$). Huygens's principle tells us that as a consequence a ray of light will be deflected in the direction in which the gravitational potential diminishes. For a ray grazing the Sun, Einstein finds a deflection of $0.83^{\prime\prime}$, and comments \cite[p.\ 108]{einsteinlorentz} that ``it would be a most desirable thing if astronomers would take up the question here raised.''\footnote{The value found by Einstein in 1911 reflects the influence of gravity on time, but does not take into account the influence of gravity on spatial geometry. The full general theory of relativity predicts a value that is twice the value predicted by the 1911 theory.}

\section{Einstein's 1916 review of general relativity}\label{GRT}   

In 1916 Einstein published a self-contained and comprehensive overview of his just-finished general theory of relativity \cite{einstein1916}, in the first part of which (``Fundamental Considerations on the Postulate of Relativity'') he pays ample attention to the conceptual foundations of the new theory. Here he also investigates the consequences of gravity for the notions of space and time, on the basis of the equivalence principle.  In particular, Einstein gives the example of a frame $K^\prime$ that rotates with respect to an inertial frame $K$ [pp.\ 115--116]\cite{einsteinlorentz}.  Concerning time in the rotating frame he writes (after having discussed the failure of Euclidean geometry)\footnote{Translation following \cite{einsteinlorentz}, but with corrections.}:
\begin{quote}
Neither can we  introduce a time in $K^\prime$ that meets the physical requirements if this time is to be indicated by clocks of identical construction at rest relatively to $K^\prime$. To see this, let us imagine two such identical clocks, placed one at the origin of the coordinates and the other at the circumference of the circle and both considered  from the ``stationary'' frame $K$. By a familiar result of the special theory of relativity, the clock at the circumference---judged from $K$---goes more slowly than the other, because the former is in motion and the other at rest. An observer at the common origin of coordinates, capable of seeing the clock at the circumference by means of light, would therefore see it lagging behind the clock beside him. As he will not make up his mind to let the velocity of light along the path in question depend explicitly on the time,  he will interpret his observations as showing that the clock at the circumference ``really'' goes more slowly than the clock at the origin. So he will be obliged to define time in such a way that the rate of a clock depends upon where the clock may be.
\end{quote}    
The structure of the argument here is the same as the one in the 1907 and 1911 articles: stationary clocks in an accelerated system, and therefore also in a system in which there is a gravitational field, will indicate a local time---but these local times do not combine into one physically reasonable global time. In the situations considered (uniform linear acceleration and uniform rotation) we should require of a physically reasonable global time that in its terms physical laws, in particular the law governing the propagation of light, should not depend explicitly on time. The latter requirement has the consequence that differences between the rates of clocks at different locations become ``objectified'': an observer who receives light signals from the clocks will be able to directly see these differences between local rates. In this way he will be able to verify that distant clocks ``really'' tick slower or faster than his own clock.  In all the given examples, both the local and the global time are supposed to be indicated by sets of clocks. In the earlier examples the global time corresponded to the indications of standard clocks in instantaneously comoving frames, in the rotating disc case global time is given by the standard clocks in the inertial frame relative to which the rotation takes place.

Of course, in the formal part of the 1916 paper things become much more abstract and general. In particular, the restriction to homogeneous gravitational fields (or special cases like fields corresponding to uniform rotation) is dropped so that, generally speaking, there will be no physically privileged global frames that provide a natural arena for the definition of clocks showing a global time. Nevertheless, it is not difficult to recognize traces of the treatment of the earlier cases, even in this part of the paper. Einstein's discussion of the gravitational redshift at the end of the 1916 review is a case in point.

In the last section of his review paper, \S 22, entitled ``Behaviour of Rods and Clocks in the Static Gravitational Field. Bending of Light-rays. Motion of the Perihelion of a Planetary orbit'', Einstein applies his new and general theory to a number of crucial cases \cite[pp.\ 160--164]{einsteinlorentz}.  The influence of gravity on clocks and the gravitational redshift are now dealt with very quickly. For a unit clock that is at rest in a static gravitational field we have for a clock period $ds = 1$ and $dx_1 =dx_2 =dx_3 = 0$.
Therefore, $g_{44} {dx_4}^2 = 1$, so that $dx_4 = 1/\sqrt{g_{44}}$.  If there is a point mass with mass $M$ at the origin of coordinates it follows from the Einstein field equations that in first approximation $g_{44} = 1 - \kappa M /4 \pi r$, with $\kappa$ the gravitational coupling constant appearing in the field equations ($\kappa = 8 \pi G/c^2$, with $G$ Newton's constant) and $r$ the radial spatial distance from the point mass.  Therefore, 
\begin{equation}\label{dx4}
dx_4 \approx 1 + \frac{\kappa M}{8 \pi r}.
\end{equation}
Einstein immediately concludes \cite[p.\ 162]{einsteinlorentz}:
\begin{quote}
Thus the clock goes more slowly if set up in the neighbourhood of ponderable masses. From this it follows that the spectral lines of light reaching us from the surface of large stars must appear displaced towards the red end of the spectrum.\footnote{Einstein subsequently shows that a light-ray grazing the Sun will be deflected by $1.7^{\prime\prime}$, twice the magnitude of the 1911 prediction, and that the orbits of the planets undergo a slow rotation, which in the case of Mercury will be $43^{\prime\prime}$ per century.}  
\end{quote} 

Einstein's reasoning here is  basically the same as in his earlier discussions of the gravitational redshift, as we shall discuss in a moment---and only this historical context of the earlier derivations makes the meaning and correctness of Einstein's derivation completely clear. Without this context misunderstandings may easily arise, as shown below.   

\section{Appraisal of the early redshift derivations }\label{crit}  

Einstein's 1916 derivation of the gravitational redshift soon became the standard one---it can still be found in general relativity textbooks. An important role in making it widely known and popular was played by the work of Eddington. Eddington was the first to make the general theory of relativity known in the English-speaking world, and his seminal publications (first of all \cite{eddington1} and \cite{eddington2}) were widely read. Of the two just-mentioned titles, especially the less technical \emph{Space, Time and Gravitation} was very influential. In this book Eddington discusses the comparison of frequencies emitted by atoms of the same kind but located at different positions, e.g.\ on the Sun and on Earth, respectively. Eddington explains the situation as follows \cite[pp.\ 128--129]{eddington2} (italics in the original):\footnote{For the sake of consistency of notation we use $g_{44}$ where Eddington wrote $\gamma$. For the comparison with Einstein's formulas it is important to note that Eddington uses units in which $\kappa/8 \pi = 1$.}
\begin{quote}
Consider an atom momentarily at rest at some point in the solar system... If $ds$ corresponds to one vibration ... we have $ds^2 = g_{44} dt^2$. The \emph{time} of vibration $dt$ is thus $1/\sqrt{g_{44}}$ times the \emph{interval} of vibration $ds$.

Accordingly, if we have two similar atoms at rest at different points in the system, the interval of vibration will be the same for both; but the time of vibration will be proportional to the inverse square-root of $g_{44}$, which differs for the two atoms. Since 
\begin{align*}
g_{44} = 1 - \frac{2M}{r}, \\
\frac{1}{\sqrt{g_{44}}} = 1 + \frac{M}{r},
\end{align*} 
very approximately.

Take an atom at the surface of the sun, and a similar atom in a terrestrial laboratory. For the first, $1 + M/r$ = 1.00000212, and for the second $1 + M/r$ is practically 1. 
The time of vibration of the solar atom is thus longer in the ratio 1.00000212, and it might be possible to test this by spectroscopic examination.

There is one important point to consider. The spectroscopic examination must take place in the terrestrial laboratory; and we have to test the period of the solar atom by the period of the waves emanating from it when they reach the earth. Will they carry the period to us unchanged? Clearly they must. The first and second pulse have to travel the same distance $r$, and they travel with the same velocity $dr/dt$; for the velocity of light in the mesh-system used is $1 - 2M/r$, and though this velocity depends on $r$, it does not depend on $t$. Hence the difference $dt$ at one end of the waves is the same as that at the other end.
\end{quote}

Eddington's account faithfully represents Einstein's 1907, 1911 and 1916 derivations. First, the atoms at different locations function as \emph{local} clocks---in modern terms, they measure \emph{proper time}, and this justifies taking $ds = 1$ as the interval for a unit period, at all locations. But second, in addition to this local time there is a global time, the ``\emph{time}'' as Eddington writes just as Einstein did in 1907 and 1911. Einstein underlined the physical importance of this global time by associating it with the indications of sets of actual clocks. Eddington does not delve into this, but he does make it clear that the global time must be used to make time comparisons between different places. This comparison is very simple: the interval $ds$ of one vibration on the Sun corresponds to a lapse $dt$ of global time at that position. Now, what is the corresponding global time lapse on Earth between the received light signals that were emitted at the beginning and the end of the atom's vibration at the sun, respectively? Because the velocity of light does not depend on $t$ (although it does depend on position), the period taken by the signals to go from the Sun to the Earth will remain the same over time, and this means that the time interval $dt$ will be transmitted unchanged. 

As we have seen in the precious sections, the latter is exactly Einstein's argument of 1907, 1911 and 1916: ``physical requirements'' tell us that global time should be introduced in such a way that physical laws, in particular the law governing the propagation of light, will not depend on time.  The immediate consequence is that global time intervals at different positions can be directly ``seen'': we do not need to make calculations about the propagation of light signals because we know that intervals of $dt$ will be transferred without change by light (or other signals). Therefore, if we want to judge at a spatial position $A$ what is the duration of a physical processes at another spatial position $B$, we can simply measure the global time interval $dt_B$ taken by the process at $B$: we know that the $B$-process will be seen at $A$ as taking up precisely the same amount of global time. This time interval can be directly compared to the time interval $dt_A$ associated with a similar process that takes place at $A$---or equivalently, both $dt_A$ and $dt_B$ can be translated into proper time intervals at $A$, which can then be compared; global and local (proper) time intervals at one position differ only by a constant factor. The frequencies of spectral lines emitted at different places can in this way be compared directly.

In an influential article from 1980, John Earman and Clark Glymour \cite{earmanglymour} criticized the early derivations of the gravitational redshift by Einstein, Eddington, and other authors who followed in their footsteps. They characterize Einstein's 1907 derivation as cumbersome, obscure and lacking clarity concerning the meanings of ``time'' and ``local time'' \cite[p.178]{earmanglymour}, without offering a detailed discussion of the article. However, in their subsequent discussion of the 1911 paper they provide a short description of the thought experiment in which radiation is emitted from $B$ to $A$ in a homogeneous gravitational field. As we have seen in some detail in section \ref{1911}, and as Earman and Glymour mention, Einstein concluded that clocks at different positions run at different rates (measured in global time), and that this implies that the velocity of light is position-dependent (as measured in global time). Earman and Glymour do not analyze Einstein's reasoning on this point, but comment \cite[pp.181--182]{earmanglymour}: 
\begin{quote}
All of the heuristic derivations of the red shift can be faulted on various technical grounds.  But to raise such objections is to miss the purpose of heuristic arguments, which is not to provide logically seamless proofs but rather to give a feel for the underlying physical mechanisms. It is precisely here that most of the heuristic red shift derivations fail---they are not good heuristics. For they are set in Newtonian or special relativistic space-time; but the red shift strongly suggests that gravitation cannot be adequately treated in a flat space-time. Einstein's resort to the notions of a variable speed of light and variable clock rates in a gravitational field can be seen as an acknowledgment, albeit unconscious, of this point; but as we will now see, these notions served to obscure the role of curvature of space-time as the light ray moves from source to receiver.
\end{quote}  

It is true that the 1907 and 1911 papers only use the (somewhat vague) principle of equivalence and can be considered faulty from the point of view of the completed general theory of relativity. In particular, there is no principled discussion of inhomogeneous fields in the 1907 and 1911 papers (although Einstein mentions in several places that he conjectures that his results for homogeneous fields will also apply to inhomogeneous ones).  But it seems unjustified to condemn Einstein's early work on this ground for being based on \emph{bad heuristics}. Moreover, and more importantly, the reproach that Einstein's treatment does not take into account the process of the transmission of light from source to receiver, and that the early derivations of Einstein, Eddington and their followers are for this reason fallacious \cite[p.\ 176]{earmanglymour} is simply incorrect---this should already be evident from our explanation of Eddington's derivation, but we shall discuss some more details below.    

After reporting on Einstein's 1916 derivation leading to Eq.\ (\ref{dx4}) and quoting Einstein's immediate conclusion (``The clock goes more slowly if set up in the neighbourhood of ponderable masses. From this it follows that the spectral lines of light reaching us from the surface of large stars must appear displaced towards the red end of the spectrum.''), Earman and Glymour continue \cite[pp.\ 182--183]{earmanglymour}:
\begin{quote}
To the modern eye, Einstein's derivation is no derivation at all, for the formula (\ref{dx4}) expresses only a co-ordinate effect, and ... Einstein provided no deduction from the theory to explain what happens to a light ray or photon as it passes through the gravitational field on its way from the Sun to the Earth. Unfortunately, Einstein's `derivation' was dressed up by he expositors of the general theory, and it quickly became codified in the literature as the official derivation.
\end{quote}
 They then explain Eddington's role in the dissemination of Einstein's error, and criticize the red shift derivation Eddington gave in his \emph{Report on the Relativity Theory of Gravitation} \cite{eddington1}: Earman and Glymour assert that the derivation confuses coordinate time with physical time and does not enter into the essential question of how the radiation emitted at the Sun is received on Earth. However, Eddington's derivation in \cite{eddington1} is essentially identical to his derivation in \cite{eddington2}, which we have reproduced above.  As should be clear from that discussion, Eddington \emph{did} take account of the role of the propagating light signal and remarked expressly on the fact that in terms of the global time $t$ the velocity of light did nor depend on time---this justified the essential point in the proof, namely that the time interval $dt$ is transferred unchanged from Sun to Earth.

It is surprising, then, to find that Earman and Glymour  present their own (admittedly flawless!) derivation of the redshift formula by arguing at length that in a static gravitational field coordinates can be chosen in such a way that the coordinate time interval is transmitted without change. Via a rather roundabout use of this premise they finally arrive at a formula that is equivalent with Einstein's \cite[pp.\ 184--185]{earmanglymour}. 

Apparently, the underlying reason of the confusion is that Earman and Glymour have looked at Einstein's (and Eddington's) formulas with all too modern eyes. In modern expositions of general relativity coordinates are considered to be purely conventional markers; in particular, the time coordinate $x_4$ does not need to have any direct physical interpretation in terms of clock indications. Formulas like (\ref{grav}) then indeed express nothing but a coordinate effect, which does not have to possess a physical significance. But Einstein did not originally approach the subject from that direction. It is true that Einstein in his definitive work on general relativity \cite{einstein1916} took a step towards the modern notion, via his insistence that frames of reference in arbitrary motion should be equivalent and that this implies that arbitrary coordinate systems can be used \cite[p.\ 776]{einstein1916},\footnote{Actually, the sense in which general relativity makes all frames of reference equivalent is controversial at least, see e.g.\ \cite{diekscov}.} but the ideas from his earlier work, in which he deemed it important to give the time coordinate always a clear concrete physical meaning, still lingered on. From this ``physical viewpoint'', and the context of the 1907 and 1911 papers, the physical properties of global time in static fields are self-evident, whereas they are in need of justification from a more modern vantage point.

\section{Conclusion: coordinates and time}\label{conclusion}

In 1921, five years after his review article on the completed general theory of relativity, Einstein gave the Stafford Little Lectures at Princeton University, in which he introduced and reviewed both the special and the general theory. The text of these lectures was published in 1922 as \emph{The Meaning of Relativity} \cite{einstein1922}. Coming back to the subject of the behavior of rods and clocks in a gravitational field \cite[pp.\ 90--92]{einstein1922}, Einstein notes that only in local inertial systems the coordinates can be chosen in such a way that they conform to ``naturally measured lengths and times''. For the case of the static field generated by a central mass, and natural coordinates usually adopted in this situation, however, a unit measuring rod will not fit exactly in a unit coordinate interval: as Einstein says, its ``coordinate length'' will be shortened---moreover, as Einstein remarks, this coordinate length, and its dependence on location and orientation, will depend on the chosen system of coordinates. So here we seem to be dealing with a coordinate effect, in the modern sense. 

However, turning to time, Einstein remarks that the interval between two beats of a unit clock ($ds=1$) corresponds to a longer ``time'' ($dx_4 > 1$) ``in the unit used in our system of coordinates'', but he immediately continues \cite[p.\ 92]{einstein1922}:
\begin{quote}
The rate of a clock is accordingly slower the greater is the mass of the ponderable mass in its neighbourhood. We therefore conclude that spectral lines which are produced on the sun's surface will be displaced towards the red, compared to the corresponding lines produced on the earth, by about $2.10^{-6}$ of their wave-lengths. 
\end{quote} This is the exact same argument as in the 1916 paper, which, as we have seen, becomes difficult to understand if we think of $dx_4$ as a completely arbitrary coordinate, and of the slowing down as a pure coordinate effect.   

Evidently, in 1921 Einstein was aware of the in principle arbitrary character of coordinates, as shown by his discussion about the behavior of measuring rods. Nevertheless, he  clung to his earlier strategy according to which the concept of time should be made as physical as possible. It had been this strategy that had helped him decisively in creating his special theory of relativity, and in taking the first steps towards general relativity. Moreover, in special cases (static gravitational field, the presence of symmetries, etc.) this same strategy is helpful even in the finished theory of general relativity. Awareness of this methodological motif that runs through Einstein's early work makes much of this work more easily understandable. This applies to the content and validity of his early relativity papers, but also to appreciating the difficulties Einstein encountered in coming to grip with relativity in its most general form, in which global time is generally no longer a sensible physical concept.


\begin{thebibliography}{99}

\bibitem{dieks}
Dieks, D., The Adolescence of Relativity:  Einstein, Minkowski and the Philosophy of Space and Time. Chapter 9 in: Petkov, V. (ed.), \emph{Minkowski Spacetime: A Hundred Years Later}, Springer, 2010. See also: Dieks, D., Time in Special Relativity. Chapter 6 in: Ashtekar, A. and Petkov, V. (eds.), \emph{The Springer Handbook of Spacetime}, Springer, 2014.
\bibitem{diekscov}
Dieks, D., Another Look at General Covariance and the Equivalence of Reference Systems,  \emph{Studies in History and Philosophy of Modern Physics}, \textbf{37}, 2006, 174--191.
\bibitem{earmanglymour}
Earman, J. and Glymour, C., The Gravitational Red Shift as a Test of General Relativity: History and Analysis, \emph{Studies in History and Philosophy of Science}, \textbf{11}, 1980, 175--214.
\bibitem{eddington1}
Eddington, A.S.,\emph{ Report on the Relativity Theory of Gravitation}. London: Fleetwood Press, 1918; 2nd edition, 1920. Reprinted by the Minkowski Institute Press, Montreal, 2014. 
\bibitem{eddington2}
Eddington, A.S., \emph{Space, Time and Gravitation}. Cambridge: Cambridge University Press, 1923. Reprinted by the Minkowski Institute Press, Montreal, 2017.
\bibitem{einstein1905}
Einstein, A.,  Zur Elektrodynamik bewegter K\"{o}rper, \emph{Annalen der Physik} \textbf{17}, 1905, 891--921. Reprinted in \cite{einsteincoll2}, pp.\ 275--310. English translation in \cite{einsteinlorentz}, pp.\ 35--65.
\bibitem{einstein1907}
Einstein, A.,  \"{U}ber das Relativit\"{a}tsprinzip und die aus demselben gezogenen Folgerungen, \emph{Jahrbuch der Radioaktivit\"{a}t und Elektronik} \textbf{4}, 1907, 411--462. Reprinted in \cite{einsteincoll2}, pp.\ 433-488. 
English translation in: Schwartz, H.M., Einstein's comprehensive 1907 essay on relativity, \emph{American Journal of Physics}, \textbf{45}, 1977, 512--517, 811--817, 899--902.
\bibitem{einstein1908}
Einstein, A., Berichtigungen zu der Arbeit ``\"{U}ber das Relativit\"{a}tsprinzip und die aus demselben gezogenen Folgerungen'',  \emph{Jahrbuch der Radioaktivit\"{a}t und Elektronik} \textbf{5}, 1908, 98--99. Reprinted in \cite{einsteincoll2}, pp.\ 493--495.
\bibitem{einstein1911}
Einstein, A., \"{U}ber den Einflu{\ss} der Schwerkraft auf die Aus\-brei\-tung des Lichtes, \emph{Annalen der Physik} \textbf{35}, 1911, 898--908. English translation in \cite{einsteinlorentz}, pp.\ 97--108. Reprinted in \cite{einsteincoll3}, pp.\ 485--497.
\bibitem{einstein1916}
Einstein, A., Die Grundlage der allgemeinen Relativit\"{a}tstheorie, \emph{Annalen der Physik}\textbf{ 49}, 1916, 769--882. Reprinted in \cite{einsteincoll6}, pp.\ 283--339. English translation in \cite{einsteinlorentz}, pp.\ 100--164.
\bibitem{einstein1922}
Einstein, A., \emph{The Meaning of Relativity}. Princeton:  Princeton University Press, 1922.
\bibitem{einsteinlorentz}
Einstein, A., Lorentz, H.A., Weyl, H., and Minkowski,, H.,  \emph{The Principle of Relativity}. New York: Dover Publications, 1952.
\bibitem{einsteincoll3}
Klein, M.J., et al. (eds), \emph{The Collected Papers of Albert Einstein, Vol.\ 3}. Princeton: Princeton University Press, 1993.
\bibitem{einsteincoll6}
Kox, A.J., et al. (eds.), \emph{The Collected Papers of Albert Einstein, Vol.\ 6}. Princeton: Princeton University Press, 1996.
\bibitem{pais}
Pais, A., \emph{Subtle is the Lord. The Science and Life of Albert Einstein.} Oxford, Oxford University Press, 1982.
\bibitem{einsteinauto}
Schilp, P.A. (ed.),  \emph{Albert Einstein, Philosopher-Scientist}. La Salle: Open Court, 1949.
\bibitem{einsteincoll2}
Stachel, J. (ed.),\emph{The Collected Papers of Albert Einstein, Vol.\ 2}. Princeton: Princeton University Press, 1989.


\end{thebibliography}
\end{document}